\documentclass[12pt,a4paper]{qts-proc-v2}
\usepackage{epsfig,cite}
\usepackage{amssymb,amsmath}
\usepackage{times}

\begin{document}
\sloppy \raggedbottom
\setcounter{page}{1}

\newpage
\setcounter{figure}{0}
\setcounter{equation}{0}
\setcounter{footnote}{0}
\setcounter{table}{0}
\setcounter{section}{0}

\newcommand{\hx}{\hat{x}}
\newcommand{\hxm}{\hat{x}^{\mu}}
\newcommand{\hxn}{\hat{x}^{\nu}}
\newcommand{\tmn}{\theta^{\mu\nu}}
\newcommand{\hD}{\hat{\Delta}}
\newcommand{\mo}{\mathcal{O}}
\newcommand{\mL}{\mathcal{L}}
\newcommand{\pd}{\partial}
\newcommand{\vp}{\varphi}
\newcommand{\bR}{\mathbb{R}}
\newcommand{\nn}{\nonumber}
\newcommand{\e}{{\rm e}}
\newcommand{\tr}{\rm Tr}
\newcommand{\del}{\delta}
\newcommand{\ra}{\rangle}
\newcommand{\la}{\langle}
\newcommand{\rar}{\rightarrow}
\newcommand{\lar}{\leftarrow}
\newcommand{\lra}{\longrightarrow}
\newcommand{\lla}{\longleftarrow}
\newcommand{\pdm}{\partial_{\mu}}
\newcommand{\pdn}{\partial_{\nu}}
\newcommand{\fmn}{F_{\mu\nu}}
\newcommand{\am}{A_{\mu}}
\newcommand{\an}{A_{\nu}}
\newcommand{\dg}{\dagger}
\newcommand{\db}{\delta_{\rm B}}
\newcommand{\bdb}{\bar{\delta}_{\rm B}}
\newcommand{\bx}{\bar{x}}
\newcommand{\fx}{x_{f}}
\newcommand{\ix}{x_{i}}
\newcommand{\vph}{\vec{\phi}}
\newcommand{\ph}{\phi}
\newcommand{\hphi}{\hat{\phi}}
\newcommand{\al}{\alpha}
\newcommand{\be}{\beta}
\newcommand{\fr}{\frac}
\newcommand{\bfx}{{\bf x}}
\newcommand{\lam}{\lambda}
\newcommand{\lag}{{\mathcal L}}
\newcommand{\cpn}{\mathbb{C}P^N}
\newcommand{\bfn}{{\bf n}}
\newcommand{\Dm}{D_{\mu}}
\newcommand{\Dn}{D_{\nu}}
\newcommand{\ep}{\epsilon}
\newcommand{\bfw}{{\bf w}}
\newcommand{\epmn}{\epsilon_{\mu\nu}}
\newcommand{\cp}{\mathbb{C}P}
\newcommand{\bz}{\bar{z}}
\newcommand{\bpd}{\bar{\pd}}
\newcommand{\sig}{\sigma}
\newcommand{\emn}{\eta_{\mu\nu}}
\newcommand{\iemn}{\eta^{\mu\nu}}
\newcommand{\mN}{\mathcal{N}}
\newcommand{\slD}{\not{\! \mbox{D}}}
\newcommand{\da}{a^{\dagger}}
\renewcommand{\th}{\theta}
\newcommand{\xmu}{x^{\mu}}
\newcommand{\xnu}{x^{\nu}}
\newcommand{\bth}{\bar{\theta}}
\newcommand{\delb}{\delta_{\rm B}}
\newcommand{\bdelb}{\bar{\delta}_{\rm B}}
\newcommand{\red}{\textcolor{red}}
\newcommand{\green}{\textcolor{green}}
\newcommand{\magenta}{\textcolor{magenta}}
\newcommand{\cyan}{\textcolor{cyan}}
\newcommand{\blue}{\textcolor{blue}}
\newcommand{\black}{\textcolor{black}}
\newcommand{\tg}{\tilde{\gamma}}
\newcommand{\s}{\scriptscriptstyle}
\newcommand{\hu}{\hat{u}}
\newcommand{\hv}{\hat{v}}
\newcommand{\I}{\bf 1}
\newcommand{\up}{\uparrow}
\newcommand{\down}{\downarrow}
\newcommand{\Rb}{\mathbb{R}}
\newcommand{\Cb}{\mathbb{C}}
\newcommand{\non}{\nonumber\\}
\newcommand{\dis}{\displaystyle}
\newcommand{\1}{\mathbb I}
\newcommand{\CC}{\mathcal C}
\newcommand{\CM}{\mathcal M}
\newcommand{\CO}{\mathcal O}
\newcommand{\CP}{\mathcal P}



\vspace*{-2cm}
\begin{flushright}
\parbox{3cm}{
KEK-TH-1053 \\ 
hep-th/0511187} 
\end{flushright}

\title{Graviton and Spherical Graviton Potentials \\
in Plane-Wave Matrix Model 
\\ \qquad - overview and perspective -}

\runningheads{H.~Shin and K.~Yoshida}{Graviton and Giant Graviton
Potentials in Plane-Wave Matrix Model}

\begin{start}


\author{Kentaroh Yoshida}{1},
\coauthor{Hyeonjoon Shin}{2}

\address{Theory Division, 
High Energy Accelerator Research Organization
(KEK),\\ Tsukuba, Ibaraki 305-0801, Japan.}{1} 

\vspace*{0.3cm}
\address{CQUeST K209, 
Sogang University, Seoul 121-742, South Korea}{2}


\begin{Abstract}
We briefly review our works for graviton and spherical graviton
potentials in a plane-wave matrix model. To compute them, it is
necessary to devise a configuration of the graviton solutions, since the
plane-wave matrix model includes mass terms and hence the gravitons are
not free particles as in the BFSS matrix model but harmonic oscillators
or rotating particles. The configuration we proposed consists of a
rotating graviton and an elliptically rotating graviton. It is applied
to the two-body interaction of spherical gravitons in the $SO(6)$
symmetric space, and then to that of point-like gravitons in the $SO(3)$
symmetric space. In both cases the leading term of the resulting
potential is $1/r^7$. This result strongly suggests that the potentials
should be closely related to the light-front eleven-dimensional
supergravity linearized around the pp-wave background.
\end{Abstract}
\end{start}


\section{Introduction}

One of the most important problems in particle physics is to clarify the
substance of M-theory which is believed as the unified theory of
superstrings. Towards the formulation of M-theory, a matrix
model approach gives a promising way. In fact, the matrix model proposed
by Banks, Fischler, Shenker and Susskind (BFSS) is conjectured to
describe a discrete light-cone quantized (light-front) M-theory
\cite{BFSS}. It is basically a one-dimensional matrix quantum mechanics
and called BFSS matrix model. It is closely related to a supermembrane
theory in eleven dimensions via the matrix regularization \cite{dWHN}.

\begin{figure}[htbp]
\centerline{\epsfig{file=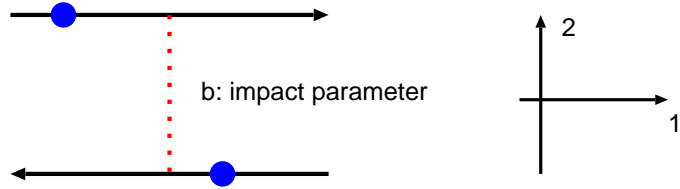,width=90mm}}
\caption{Two graviton scattering in the BFSS case}\label{BFSS:fig}
\end{figure}

M-theory is considered to contain the eleven-dimensional supergravity as
a low energy effective theory. On the other hand, the matrix model is
shown to contain the gravity and therefore this fact gives a strong
evidence for the conjecture. More concretely speaking, graviton
potentials in the light-front eleven-dimensional supergravity can be
reproduced from the computation in the matrix model. For example, let us
consider two graviton scattering in the BFSS matrix model. A single
graviton is described by $1\times 1$ matrix,
\begin{equation}
\label{graviton}
 X^I = a^I + v^I t \quad \mbox{(free~particle)}\,,  
\end{equation}
which is a classical solution. 
The configuration for two-graviton scattering is given by \begin{eqnarray}
B^1 = \frac{1}{2}\begin{pmatrix}
vt & 0 \\ 0 & -vt
\end{pmatrix}
\,, \quad B^2 = \frac{1}{2}\begin{pmatrix}
b & 0 \\ 0 & -b
\end{pmatrix}
\,, \quad B^i=0 \quad (i>2)\,, 
\label{eq2}
\end{eqnarray}
which is drawn in Fig.\,\ref{BFSS:fig}. 
By using the background field method and integrating out the
fluctuations around the configuration (\ref{eq2}), the potential is computed 
as a function of the impact
parameter $b$\,. The resulting potential is given by
\begin{eqnarray}
V(b) = - \frac{15}{16}\frac{v^4}{b^7} + \mathcal{O}
\left(\frac{v^6}{b^{11}}\right)\,. 
\end{eqnarray} 
The term proportional to $b$ and 
the lower order terms, that is, $1/b$\,, $1/b^3$ and $1/b^5$, are 
canceled out basically because of supersymmetries. The term with 
$1/b^9$ also does not appear and hence the subleading term is $1/b^{11}$. 

The one-loop result in the matrix model 
agrees with the potential derived by evaluating a tree diagram 
in the light-front eleven-dimensional supergravity, including the numerical
factors. We would like to note that, in fact, the one-loop result has
turned out to be exact entirely due to the power of 16 supersymmetries
\cite{PSSHKS}.
Spherical membrane scattering is also discussed in
\cite{KT}. For the detail, see \cite{review}. 

From now on, we will overview graviton and spherical graviton potentials
in a plane-wave matrix model (PWMM) \cite{BMN}. This matrix model may be
considered as a generalization of the BFSS matrix model to the pp-wave
background which has non-vanishing curvature. It is very interesting
to consider whether the matrix model on the pp-wave background can
describe the gravity correctly or not.

\section{Plane-Wave Matrix Model} 

Here we shall briefly introduce a plane-wave matrix model. 
The action of the matrix model is given by \cite{BMN}
\begin{eqnarray}
&& S_{\rm pp} = \int\!\! dt\,
\mathrm{Tr} \Bigl[ \frac{1}{2R} D_t X^I D_t X^I + \frac{R}{4} ( [
X^I, X^J] )^2 + i \Theta^\dagger D_t \Theta  - R \Theta^\dagger 
\gamma^I [\Theta, X^I ] 
\nonumber \\
&& \qquad \qquad 
-\frac{1}{2R} \left( \frac{\mu}{3} \right)^2 (X^i)^2
-\frac{1}{2R} \left( \frac{\mu}{6} \right)^2 (X^a)^2 
- i \frac{\mu}{3}
\epsilon^{ijk} X^i X^j X^k - i \frac{\mu}{4} \Theta^\dagger \gamma^{123}
\Theta \Bigr]\,,
\end{eqnarray}
where the indices of the transverse nine-dimensional space are
$I,J=1,\ldots,9$ and $R$ is the radius of the circle compactified along
$x^-$\,. All degrees of freedom are $N\times N$ Hermitian matrices and
the covariant derivative $D_t$ with the gauge field $A$ is defined by
$D_t = \partial_t-i[A,~~]$\,. The plane-wave matrix model can be
obtained from the supermembrane theory on the pp-wave background
\cite{DSR,SY} via the matrix regularization \cite{dWHN}.  In particular,
in the case of the pp-wave, the correspondence of superalgebra between
the supermembrane theory and the matrix model, including brane charges,
is established by the works \cite{SY} and \cite{HS1}.

This matrix model may be considered as a deformation of the BFSS matrix
model while it still preserves linearly realized 16 supersymmetries. 
The plane-wave matrix
model allows a static 1/2 BPS fuzzy sphere with zero light-cone energy
to exist as a classical solution, since the action of the matrix model
includes the Myers term \cite{Myers}. The structure of the vacua is
enriched with the fuzzy sphere. The spectra around the vacua are now
fully clarified \cite{DSR,DSR2}. The trivial vacuum $X^I=0$ has also
been identified with a single spherical five-brane vacuum \cite{TM5}.
Except for the static fuzzy sphere,there are various classical solutions
and those have been well studied (e.g., see \cite{Bak}). 
BPS properties of fuzzy sphere have been investigated  
in several papers \cite{DSR,SY3,HSKY}.  Thermal
stabilities of classical solutions have also been investigated in
\cite{Furuuchi}.  

Our purpose here is to compute the graviton potential in the plane-wave
matrix model. One should note that the graviton solution
(\ref{graviton}) as a free particle is not a classical solution any
more, since the plane-wave matrix model contains mass terms in contrast
to the BFSS matrix model. The graviton solution in the plane-wave matrix
model is represented by a harmonic oscillator or a rotating particle.
Then it is necessary to devise the setup to examine the two-graviton
scattering in the plane-wave matrix model. In the next section we will
discuss the configuration of the gravitons for the computation of the
potential. Before going to the explanation of the setup, in the next
subsection we will explain the background field method and the exactness
of the one-loop calculation in the $\mu\to\infty$ limit.

\subsection{Background Field Method and One-Loop Exactness}

To compute the interaction potential we use the background field method
as usual. To begin with, the matrix variables are decomposed into the
background and the fluctuations as
\begin{eqnarray}
\label{cl+qu}
 X^I = B^I + Y^I\,, \quad \Theta = 0 + \Psi\,. 
\end{eqnarray}
where $B^I$ are the classical background fields while $Y^I$ and 
$\Psi$ are the quantum fluctuations around them. 
Here the fermionic background is set to zero. 

In order to perform the path integration, we take the background field
gauge which is usually chosen in the matrix model calculation as
\begin{equation}
D_\mu^{\rm bg} A^\mu_{\rm qu} \equiv 
D_t A + i [ B^I, X^I ] = 0 ~.
\label{bg-gauge}
\end{equation}
Then the corresponding gauge-fixing $S_\mathrm{GF}$ and Faddeev-Popov
ghost $S_\mathrm{FP}$ terms are given by
\begin{equation}
S_\mathrm{GF} + S_\mathrm{FP} = \int\!dt \,{\rm Tr} \left( - \frac{1}{2}
(D_\mu^{\rm bg} A^\mu_{\rm qu} )^2 - \bar{C} \partial_{t} D_t C + [B^I,
\bar{C}] [X^I,\,C] \right)\,.  \label{gf-fp}
\end{equation}
Now by inserting the decomposition of the matrix fields (\ref{cl+qu})
into the matrix model action, we get the gauge fixed plane-wave action
$S$ $(\equiv S_{\rm pp} + S_\mathrm{GF} + S_\mathrm{FP})$ expanded around
the background.  The resulting action is read as 
$S =  S_0 + S_2 + S_3 + S_4$\,,  
where $S_n$ represents the action of order $n$ with respect to the
quantum fluctuations. 
Here we write down only the second order part: 
\begin{eqnarray}
&& \hspace*{-0.6cm}
S_2 = \int\!\! dt\,\mathrm{Tr} \bigg[ \,
      \frac{1}{2} ( \dot{Y}^I)^2 - 2i \dot{B}^I [A, \, Y^I] 
        + [B^I , \, B^J] [Y^I , \, Y^J] \nn \\ 
&&   + \frac{1}{2}([B^I , \, Y^J])^2      
- i \mu \epsilon^{ijk} B^i Y^j Y^k 
- \frac{1}{2} \left( \frac{\mu}{3} \right)^2 (Y^i)^2 
        - \frac{1}{2} \left( \frac{\mu}{6} \right)^2 (Y^a)^2 
        \nn \\ 
&&         + i \Psi^\dagger \dot{\Psi} 
        -  \Psi^\dagger \gamma^I [ \Psi , \, B^I ] 
        -i \frac{\mu}{4} \Psi^\dagger \gamma^{123} \Psi  
\nn \\ 
& &      - \frac{1}{2} \dot{A}^2  - \frac{1}{2} ( [B^I , \, A])^2 
        + \dot{\bar{C}} \dot{C} 
        + [B^I , \, \bar{C} ] [ B^I ,\, C] \,
     \bigg]\,.
\label{bgaction} 
\end{eqnarray} 
The above expression is given in the Minkowski formulation and hereafter
we will not move to the Euclidean formulation. 
The first order part $S_1$ vanishes by using the equation of motions. 
The zeroth order part is also zero for all of the classical
configurations we consider later, although the solutions are rotating.

For the justification of one-loop computation or the semi-classical
analysis, it should be made clear that $S_3$ and $S_4$ 
can be regarded as perturbations.  For this
purpose, following \cite{DSR}, we rescale the fluctuations and
parameters as
\begin{eqnarray}
A   \rightarrow \mu^{-1/2} A   ~,~~~
Y^I \rightarrow \mu^{-1/2} Y^I ~,~~~
C  \rightarrow \mu^{-1/2} C   ~,~~~
t \rightarrow \mu^{-1} t ~.
\label{rescale}
\end{eqnarray}
Under this rescaling, the action $S$ in the fuzzy sphere background becomes
\begin{align}
S =  S_2 + \mu^{-3/2} S_3 + \mu^{-3} S_4 ~,
\label{ssss}
\end{align}
where the parameter $\mu$ in $S_2$, $S_3$ and $S_4$ has been replaced by
1 and so those do not have $\mu$ dependence. Now it is obvious that, in
the large $\mu$ limit, $S_3$ and $S_4$ can be treated as perturbations
and the one-loop computation gives the sensible result. Note that the
analysis in the $S_2$ part is exact in the $\mu\rightarrow \infty$
limit. We can calculate the exact spectra around an $N$-dimensional
irreducible fuzzy sphere in the $\mu \rightarrow \infty$ limit, by
following the method in the work \cite{DSR} (For the detail of the
calculation, see \cite{DSR,HSKY}). The exact spectra are useful to 
compute the giant graviton potential.

\section{Giant Graviton Scattering} 

Here we consider a two-body scattering of spherical gravitons (fuzzy
spheres) \cite{HSKY,HSKY-potential} which expand in the $SO(3)$
symmetric space. These solutions are considered as giant gravitons. 
Hence we call the potential between the spherical gravitons the giant
graviton potential. 

\subsection{One-Loop Flatness}

\begin{figure}[htbp]
\centerline{\epsfig{file=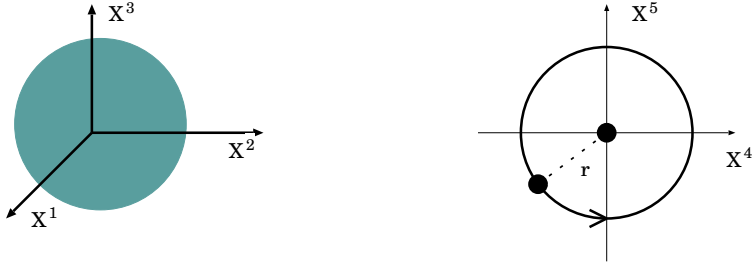,width=100mm}}
\caption{BPS configuration of two spherical gravitons in the PWMM case}\label{flat-tra:fig}
\end{figure}

As a first trial, we proposed a setup of the giant gravitons to compute the
interaction potential, drawn in Fig.\,\ref{flat-tra:fig}. 

Two fuzzy spheres expand in the $SO(3)$ symmetric space, and in a sub-plane
in the $SO(6)$ symmetric space a spherical membrane (with $p^+=N_1/R$)
is sitting at the origin of the $SO(6)$ symmetric space and the other
one (with $p^+=N_2/R$) rotates with $r$\,.

The background is described as 
\begin{eqnarray}
&& B^{I} = \begin{pmatrix}
B^I_{(1)} & 0 \\
 0  &  B^I_{(2)}
\end{pmatrix}
, \quad \begin{array}{c}
B^I_{(1)}: N_1\times N_1 \\
B^I_{(2)}: N_2\times N_2
\end{array}
, \quad
N = N_1 + N_2 \nn \vspace*{0.2cm} \\
&& B^i_{(s)} = \frac{\mu}{3}J^i_{(s)}, \quad
B^6_{(s)} = \dots = B^9_{(s)} = 0 \qquad (s=1,2)\,. \nn
\end{eqnarray} 
The rotation of the first fuzzy sphere is rotating around the origin 
with a constant radius $r$\,,  
\begin{eqnarray}
&& B^4_{(1)} = r \cos\left(\frac{\mu}{6}t\right){\bf
 1}_{N_1\times N_1}, \quad
B^5_{(1)} = r\sin\left(\frac{\mu}{6}t\right)
{\bf 1}_{N_1\times N_1}\,,  \nn 
\end{eqnarray}
and the second fuzzy sphere is sitting at the origin. 
When we rescale the variables as in (\ref{rescale})\,, the parameter
$r$ is also rescaled as $r \to \mu r$\,.  

For this setup, we have computed the potential by integrating out the
fluctuations around this background. The resulting potential is,
however, zero and hence the system is shown to be BPS. Nevertheless, the
system is still important as a start point since we can expect to obtain
non-trivial potential by breaking the remaining supersymmetries. Thus
the next task is to consider how to break the remaining
supersymmetries. 

\subsection{Elliptic Deformation}

In order to break the remaining supersymmetries, we hit on an
elliptic deformation of the setup in Fig.\,\ref{flat-tra:fig}.  
That is, we considered a configuration in a sub-plane in the
$SO(6)$ symmetric space where a spherical membrane (with $p^+=N_1/R$)
rotates with a constant radius $r_1$ and another one (with $p^+=N_2/R$)
elliptically rotates with $r_2 \pm \epsilon$\,. 
The motion of the second fuzzy sphere is elliptically deformed with the
infinitesimal parameter $\epsilon$\,. This parameter plays the similar 
role with the velocity $v$ of the graviton in the BFSS case where $v$ is
also assumed to be sufficiently small.  

The motion of the first and the second fuzzy spheres are represented by,
respectively,
\begin{eqnarray}
&& B^4_{(1)} = r_1 \cos\left(\frac{\mu}{6}t\right){\bf
 1}_{N_1\times N_1}, \quad
B^5_{(1)} = r_1\sin\left(\frac{\mu}{6}t\right)
{\bf 1}_{N_1\times N_1}  \nn \\ 
&& B^4_{(2)} = (r_2 +\epsilon) \cos\left(\frac{\mu}{6}t\right)
{\bf 1}_{N_2\times N_2}, \quad
B^5_{(2)} = (r_2 - \epsilon)\sin\left(\frac{\mu}{6}t\right)
{\bf 1}_{N_2\times N_2}  \nn
\end{eqnarray}
The fuzzy spheres are expanding in the $SO(3)$ symmetric space. 
The parameters $r_{1,2}$ and $\epsilon$ are also rescaled as 
$r_{1,2}\,,~\epsilon \to \mu r_{1,2}\,,~\mu\epsilon$\,. 

For this setup we have computed the effective action by using the
background field method. The resulting effective action with respect to
$r \equiv r_2 - r_1$ is\footnote{In fact, $r$ should be
regarded as $|r|$\,.}
\begin{eqnarray}
&& \hspace*{-1cm}
\Gamma_{\rm eff} = \epsilon^4 \int\!\!dt\, \biggl[
 \frac{35}{2^7 \cdot 3} \frac{N_1 N_2}{r^7} -\frac{385}{2^{11} \cdot 3^3}
  \big[ 2 (N_1^2 + N_2^2) -1 \big] \frac{N_1 N_2}{r^9} 
+  \mathcal{O} \left(\frac{1}{r^{11}} \right) \biggr]
+ \mathcal{O} (\epsilon^6)\,.  
\end{eqnarray}
This result strongly suggests that the spherical membranes should be
interpreted as spherical gravitons as discussed by Kabat and Taylor
\cite{KT}. Here we should remark that the subleading term is $1/r^9$ and
it is repulsive. In the BFSS case the subleading term is $1/r^{11}$
order and it implies the dipole-dipole interaction. According to the
interpretation, the $1/r^9$ term would imply the dipole-graviton
interaction. This is a new effect intrinsic to the pp-wave background.

\begin{figure}[htbp]
\vspace*{0.3cm}
\centerline{\epsfig{file=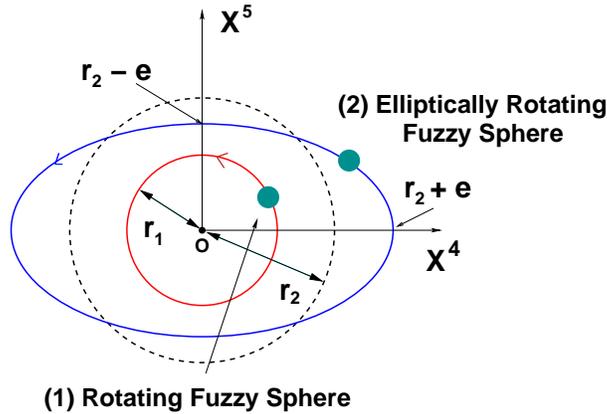,width=80mm}} \caption{Scattering of
two spherical gravitons in the PWMM case}\label{ellipsfig}
\end{figure}

\section{Point-Like Graviton Scattering}

\begin{figure}[htbp]
\vspace*{0.3cm}
\centerline{\epsfig{file=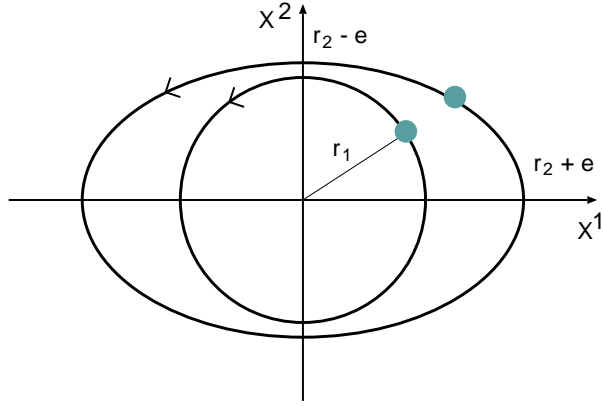,width=80mm}}
\caption{Two point-like gravitons in the PWMM case}\label{gravs:fig}
\end{figure}

Now we will discuss a two-body scattering in the $SO(3)$ symmetric
space \cite{HSKY-graviton}. Then the configuration for the computation
consists of two {\it point-like} gravitons in contrast to the spherical
membrane cases, since fuzzy spheres cannot expand due to the directions
of the constant flux. 
This background is represented by
\begin{eqnarray}
&& B^I = \begin{pmatrix} 
B^I_{(1)}  &  0 \\ 0 & B^I_{(2)} 
\end{pmatrix}
 \qquad (I=1,\ldots,9)\,, \nn \\
&& B^{1}_{(1)} = r_1\cos\left(\frac{\mu}{3}t\right)\,, \qquad 
B^{2}_{(1)} = r_1 \sin\left(\frac{\mu}{3}t\right)\,, \nn \\ 
&& B^{1}_{(2)} = (r_2+\epsilon)\cos\left(\frac{\mu}{3}t\right)\,, \qquad  
B^2_{(2)} = (r_2-\epsilon)\sin\left(\frac{\mu}{3}t\right) \nn \\
&& B^3_{(s)} = B^a_{(s)} = 0 \qquad (s=1,2~;~a=4,\ldots,9)\,. 
\end{eqnarray} 
In this setup two point-like gravitons are rotating in the 1-2 plane. 
One of them is rotating with a constant radius $r_1$ 
and the other is elliptically rotating with $r_2\pm\epsilon$\,,   
as depicted in Fig.\,\ref{gravs:fig}\,. 

We can compute the potential as a function of $r\equiv r_2-r_1$\,, and 
the resulting potential is given by
\begin{align}
\Gamma_{\rm eff} &=
\epsilon^4\int dt \left[~\frac{35}{24} \frac{1}{r^7}
+ \frac{385}{576} \frac{1}{r^9} + \mathcal{O}\left(\frac{1}{r^{11}}
\right)~\right] + \mathcal{O}(\epsilon^6)\,. \nn
\end{align}
The leading term is also $1/r^7$\,, but the subleading term is 
attractive in contrast to the spherical membrane cases. 
The numerical coefficients are also different from the case 
in the $SO(6)$ symmetric space, but it is not suspicious since 
the transverse $SO(9)$ symmetry is not preserved any more and 
it is broken to $SO(3)\times SO(6)$\,.

\section{Conclusion and Discussion}

We have discussed two-body scatterings of gravitons and spherical
gravitons in the plane-wave matrix model. The resulting potentials in
both cases have $1/r^7$ term as the leading term.  The $1/r^7$ behavior
strongly suggests that the potentials should be related to the
light-front eleven-dimensional supergravity. Eventually, it should be
the linearized supergravity around the pp-wave background. It is an
interesting direction to find the corresponding configuration in the
supergravity side i.e., the tree diagram leading to the potential
obtained in the matrix model computation. In the supergravity analysis
it is necessary to take asymptotic states, but the spectrum of the
linearized supergravity around the pp-wave background has already been
obtained in \cite{Kimura}. By using this spectrum, it would be possible
to rederive the potential from the supergravity, including the numerical
coefficients as in the BFSS case. The work \cite{LMW} would be helpful
to study in this direction. We hope that we could report on the subject
in another place in the near future.

It is worth noting again the subleading terms. The subleading term in
the plane-wave matrix model case is $1/r^9$ in comparison to $1/r^{11}$
in the BFSS matrix model case. As for the signature of the term, it is
repulsive in the $SO(6)$ case and attractive in the $SO(3)$ case. In the
BFSS case the $1/r^{11}$ term is interpreted as a dipole-dipole
interaction. According to an analogy to the BFSS case, the $1/r^9$ term
should be interpreted as a graviton-dipole interaction. This term is
intrinsic to the pp-wave background and may lead to some new physics.
Thus it is valuable to clarify the meaning of the subleading term in
connection with the geometry of the pp-wave background. 

We hope that our potentials would be an important clue to 
clarify some features of M-theory on the pp-wave background, 
and furthermore that they shed light on M-theory on curved backgrounds.

%


\section*{Acknowledgments}

The work of K.~Y.\ is supported in part by JSPS
Research Fellowships for Young Scientists. 
The work of H.S. was supported by grant No. R01-2004-000-10651-0
from the Basic Research Program of the Korea Science and
Engineering Foundation (KOSEF).


\end{document}